\documentclass[aps,pra,twocolumn,twoside,a4paper,10pt,amsmath,amssymb,superscriptaddress]{revtex4-1}

\usepackage{mathrsfs}
\usepackage{graphicx}
\usepackage{amsmath}
\usepackage{amssymb}
\usepackage{amsfonts}
\usepackage{color}
\usepackage{CJK}
\usepackage{leftidx}
\def\bra#1{\left\langle{#1}\right|}
\def\ket#1{\left|{#1}\right\rangle}
\def\braket#1#2{\left\langle{{#1}}\mathrel{\left|{\vphantom{{#1}{#2}}}\right.\kern-\nulldelimiterspace}{{#2}}\right\rangle}

\begin{document}
\title{Heisenberg-limited Frequency Estimation via Driving through Quantum Phase Transitions}
\author{Min Zhuang}
\affiliation{Guangdong Provincial Key Laboratory of Quantum Metrology and Sensing $\&$ School of Physics and Astronomy, Sun Yat-Sen University (Zhuhai Campus), Zhuhai 519082, China}

\author{Hongtao Huo}
\affiliation{Guangdong Provincial Key Laboratory of Quantum Metrology and Sensing $\&$ School of Physics and Astronomy, Sun Yat-Sen University (Zhuhai Campus), Zhuhai 519082, China}

\author{Yuxiang Qiu}
\affiliation{Guangdong Provincial Key Laboratory of Quantum Metrology and Sensing $\&$ School of Physics and Astronomy, Sun Yat-Sen University (Zhuhai Campus), Zhuhai 519082, China}
\affiliation{State Key Laboratory of Optoelectronic Materials and Technologies, Sun Yat-Sen University (Guangzhou Campus), Guangzhou 510275, China}

\author{Wenjie Liu}
\affiliation{Guangdong Provincial Key Laboratory of Quantum Metrology and Sensing $\&$ School of Physics and Astronomy, Sun Yat-Sen University (Zhuhai Campus), Zhuhai 519082, China}

\author{Jiahao Huang}
\altaffiliation{Email: hjiahao@mail2.sysu.edu.cn, eqjiahao@gmail.com}
\affiliation{Guangdong Provincial Key Laboratory of Quantum Metrology and Sensing $\&$ School of Physics and Astronomy, Sun Yat-Sen University (Zhuhai Campus), Zhuhai 519082, China}

\author{Chaohong Lee}
\altaffiliation{Email: lichaoh2@mail.sysu.edu.cn, chleecn@gmail.com}
\affiliation{Guangdong Provincial Key Laboratory of Quantum Metrology and Sensing $\&$ School of Physics and Astronomy, Sun Yat-Sen University (Zhuhai Campus), Zhuhai 519082, China}
\affiliation{State Key Laboratory of Optoelectronic Materials and Technologies, Sun Yat-Sen University (Guangzhou Campus), Guangzhou 510275, China}

\begin{abstract}
  High-precision frequency estimation is an ubiquitous issue in fundamental physics and a critical task in spectroscopy.
  Here, we propose a quantum Ramsey interferometry to realize high-precision frequency estimation in spin-$1$ Bose-Einstein condensate via driving the system through quantum phase transitions(QPTs).
  In our scheme, we combine adiabatically driving the system through QPTs with $\frac{\pi}{2}$ pulse to realize the initialization and recombination.
  Through adjusting the laser frequency under fixed evolution time, one can extract the transition frequency via the lock-in point.
  The lock-in point can be determined from the pattern of the population measurement.
  In particular, we find the measurement precision of frequency can approach to the Heisenberg-limited scaling.
  Moreover, the scheme is robust against detection noise and non-adiabatic effect.
  Our proposed scheme does not require single-particle resolved detection and is within the reach of current experiment techniques.
  Our study may point out a new way for high-precision frequency estimation.
\end{abstract}
\date{\today}

\maketitle

\section{Introduction\label{Sec1}}
%
The high-precision frequency estimation is important for many areas ranging from fundamental physics and modern metrology science to molecular spectroscopy and global position systems~\cite{Kleppner2006,Chou2010,Ashby1999,Nature506,RMP851103,RMP851083,prl104070802,prl109203002,MSGrewal2013}.
%
%
The history of precision spectroscopy with the atomic and molecular beam resonance method started from $1930$s, which was originally proposed by I. I. Rabi~\cite{IIRabi1937}.
By scanning the frequency of the electromagnetic excitation around the exact resonance, a symmetric measurement signal with respect to the resonant point can be observed.
The symmetric measurement signal can be used as the frequency lock-in signal and one can determine the value of frequency from it.
Further, to improve the measurement precision of frequency, Ramsey technique of separated oscillating field was proposed ~\cite{NFRamsey1950,NFRamsey1963} and has been widely applied in experiments~\cite{JLHall2006,TWHansch2006,NHinkley2013,MTakamoto2003,TSteinmetz2008,MSGrewal2013,HMargolis2014}.

In Atomic, Molecular and Optical (AMO) systems, Ramsey interferometry is a generalized tool for frequency estimation.
For $N$ two-level atoms with internal states $\ket{e}$(excited state) and $\ket{g}$(ground state),
the transition frequency between the two internal states is $\omega=(E_{e}-E_{g})/\hbar$ (we set $\hbar=1$ in the following).
In the conventional Ramsey interferometry, assume the input state is a product state $\ket{\Psi}_\textrm{{Pro}}=[(\ket{e}+\ket{g})/\sqrt{2}]^{\bigotimes N}$, which can be prepared by a $\frac{\pi}{2}$ pulse with laser frequency $\omega_L$ slightly detuned from the atomic transition frequency $\omega$.
Then each atom undergoes a free evolution of duration $T$ and a second $\frac{\pi}{2}$ pulse is applied.
Finally, a measurement of the atomic state is performed.
During the evolution time $T$ the atoms gather up a relative phase $\phi=\delta T$ with detuning $\delta=\omega-\omega_{L}$, which can be estimated from the measurement data.
Due to the frequency $\omega_L$ and the evolution time $T$ are known, one can recover the transition frequency $\omega$ from the relative phase $\phi$.
Ideally, using the product state $\ket{\Psi}_\textrm{{Pro}}$ as input, the measurement precision can achieve the Standard quantum limit(SQL), i.e., $\Delta \omega= 1/\sqrt{N}T$~\cite{MayEKim2015,RSarkar2015,DJWineland1992,DJWineland1994}, which had been realized in atomic clocks~\cite{Santarelli1999,Wilpers2002,Ludlow2008}.
%
%

It is well known that quantum entanglement is a useful resource for improving the measurement precision over the SQL.
The metrologically useful many-body quantum entangled states include spin squeezed states~\cite{JMKitagawa1991,JMKitagawa1993,PBouyer1997,VMeyer2001,ALouchetChauvet2010,JGBohnet2016}, spin cat state~\cite{J.Huang2015}, twin Fock (TF) state ~\cite{PRA023810}, Greenberger-Horne-Zeilinger (GHZ) state~\cite{JJBollinger1996} and so on.
Thus, a lot of endeavors had been made to generate various kinds of entangled input states.
In general, one can prepare the desired entangled state via dynamical evolution~\cite{MFRiedel2010,BLucke2011,Bookjans2011,Strobel2014,Gabbrielli163002} or adiabatic driving~\cite{CLee2006,CLee2009,PRL1200632012018,PRA930436152016,ZZhang2013,J.Huang2015,J.Huang2018}.
For GHZ state, the measurement precision of frequency can improve to the Heisenberg limit, i.e., $\Delta\omega=1/{NT}$~\cite{JJBollinger1996,VMeyer2000,DLeibfried2003,DLeibfried2004,DLeibfried2005,IDLeroux2010,CGross2010,MFRiedel2010,BLucke2011,TMonz2011,WMuessel2015,SRavid2018}.
However, it is hard to prepare the GHZ state with large atomic number in experiments.
For TF state, the measurement precision of frequency can beat the SQL, i.e., $\Delta \omega= 1/{\sqrt{N({N/2+1})}T}$ ~\cite{PRA023810}.
Moreover, the TF state has been generated deterministically by adiabatic driving in spin-$1$ atomic Bose-Einstein condensate with more than $1000$ atoms~\cite{ZZhang2013,XLuo2017,SGuo2021}.

However, to realize quantum-enhanced frequency estimation via entanglement, single-particle resolved detection is assumed to be necessary~\cite{PRA023810,DJWineland1992,SFHuelga1997,EDavis2016,FFrowis2016,TMacr2016,DLinnemann2016,OHosten2016,SSSzigeti2017,SPNolan2017,JHuang2014}, which has been a bottleneck in practical experiments.
Moreover, imperfect initial state preparation, imperfect recombination and detection both are the key obstacles that limit the improvement of measurement precision via many-body entanglement.
Based on the quantum-enhanced frequency estimation via TF state, it is natural to ask:(i) Can one achieve Heisenberg-limited frequency measurement without single-particle resolved detection?
(ii)If the Heisenberg-limited measurements are available, what are the influences of imperfection on frequency estimation in practical experiment?

In this article, we propose a scheme to implement Heisenberg-limited frequency measurement via driving through quantum phase transitions without single-particle resolved detection.
Our scheme is based on ferromagnetic spin-$1$ Bose-Einstein condensate under an external magnetic field.
Through adjusting the laser frequency, we can extract the transition frequency $\omega$ according to the frequency lock-in signal.
For every fixed laser frequency $\omega_{L}$, one can implement quantum interferometry for frequency estimation.
The interferometry consists of four steps: (a)initialization, (b)interrogation, (c) recombination, and (d) measurement.
In the interferometry , we combine adiabatically driving through QPTs with $\frac{\pi}{2}$ pulse to realize the initialization and recombination.
We find that the population measurement is symmetric respect to detuning $\delta=\omega-\omega_{L}$ and reachs its maximum at the lock-in point $\omega=\omega_{L}$.
Thus, population measurement can be the frequency lock-in signal and one can obtain the value of frequency from it.
Especially, we find the measurement precision of the frequency can approach the Heisenberg-limited scaling.
Compared with conventional proposal of quantum-enhanced frequency estimation via parity measurement~\cite{PRA023810}, our scheme does not require single-particle resolved detection.
%
%

Further, we study the robustness of our scheme against detection noise and non-adiabatic effect.
We find that the detection noise and non-adiabatic effect do not induce any frequency shift on the frequency lock-in signals.
For detection noise, the measurement precision of frequency $\Delta\omega$ can beat the SQL when $\sigma\leq 0.7\sqrt{N}$ in our consideration.
For non-adiabatic effect, the measurement precision of frequency $\Delta\omega$ can still beat the SQL when the sweeping rate $\beta$ is moderate.
Our proposed scheme may open up a feasible way of measuring frequency at the Heisenberg limit without single-particle resolved detection.

The paper is organised as follows.
In Sec.~\ref{Sec2}, we introduce our scheme for frequency estimation.
In Sec.~\ref{Sec3}, within our scheme, we study the frequency lock-in signals and the frequency measurement precision in detail.
In Sec.~\ref{Sec4}, the robustness to practical detection noise and the influences of non-adiabatic effect are discussed.
In Sec.~\ref{Sec5}, a brief summary is given.
\section{general scheme \label{Sec2}}
\begin{figure}[!htp]
 \includegraphics[width=\columnwidth]{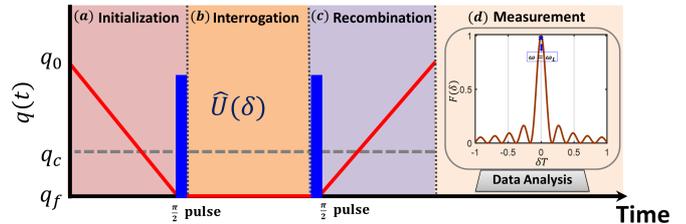}
  \caption{\label{Fig1}(color online).
  The proposal of frequency measurement scheme via driving through QPTs in spin-$1$ BEC.
  The scheme consists of four steps: initialization, interrogation, recombination, and measurement.
  (a)In the initialization stage, an initial state is prepared. Then, sweeping $q(t)$ from $q_{0}$ towards $q_{f}$ very slowly to generate an entangled state and then a $\pi/2$-pulse is applied on it to prepare the input state.
  (b)In the interrogation step, the input state undergoes an interrogation stage for signal accumulation with $\phi=\delta T$.
  (c)In the recombination step, a second $\pi/2$-pulse is applied and then sweeping $q(t)$ from $q_{f}$ towards $q_{0}$ very slowly for recombination.
  (d)In the measurement step, applying a suitable observable measurement on the final state.
  The pattern of observable measurement results versus $\delta$ can tell us the frequency lock-in point $\omega=\omega_{L}$.
  Here, $c_{2}=-1$, $q_0 =-q_f=3$, $T=100$ and $ \beta = 0.01$.
  }
\end{figure}
Our proposal for frequency estimation via driving through quantum phase transitions is presented below.
We consider an ensemble of spin-$1$ atoms with three Zeeman levels: $\ket{F=1,m=1}$, $\ket{F=1,m=0}$, $\ket{F=1,m=-1}$.
Throughout this paper, we assume all time-evolution processes are unitary and abbreviate the three Zeeman levels to $\ket{1,1}$, $\ket{1,0}$ and $\ket{1,-1}$ respectively.
We choose state $\ket{1,1}$ as the excited state $\ket{e}$ and state $\ket{1,-1}$ as the ground states $\ket{g}$, our goal is to measure the transition frequency $\omega$ between the two states $\ket{1,1}$ and $\ket{1,-1}$.
The system states can be represented in terms of the Fock basis $\ket{N_{-1},N_{0},N_{1}}$.
Here, $\hat{N}_{m}=\hat{a}^{\dagger}_{m}\hat{a}_{m}$ denotes the particle number operator of atoms in $\ket{1,m}$, with the creation operator $\hat{a}^{\dagger}_{m}$ and the annihilation operator $\hat{a}_{m}$.

In our scheme, by adjusting the laser frequency $\omega_{L}$, we can extract the transition frequency $\omega$ according to the frequency lock-in signals.
For every fixed laser frequency $\omega_{L}$, one can implement quantum interferometry for frequency estimation.
The interferometry consists of four steps: (a) initialization, (b) interrogation, (c) recombination, (d) measurement, as shown in Fig.~\ref{Fig1}.
Now, we introduce the four steps in detail.
In the initialization step, we consider the initial state with all $N$ atoms in the state $\ket{1,0}$ and the total atomic number $N$ being an even integer.
The evolution of the initial state $\ket{\Psi}_\textrm{in}=\ket{0,N,0}$ is governed by the following Hamiltonian:
\begin{eqnarray}\label{Eq:Hamitonian_QPT}
  \hat{H}_{\textrm{QPT}}=\frac{c_2}{2N}[2(\hat{a}^{\dagger}_{0}\hat{a}^{\dagger}_{0}\hat{a}_{1}\hat{a}_{-1}
  +\hat{a}_{0}\hat{a}_{0}\hat{a}^{\dagger}_{1}\hat{a}^{\dagger}_{-1})\nonumber\\
         +(2\hat{N_0}-1)(N-\hat{N_0})]-q(t)\hat{N_0}.
\end{eqnarray}
Here, $|c_2|(c_2<0)$ describes the rate of spin mixing process, $q=(\varepsilon_{+1}+\varepsilon_{-1})/2-\varepsilon_{0}$, with $\varepsilon_{m}$ being the energy of the state $\ket{1, m}$,
and $q(t)$ can be tuned linearly with time in experiment.
The system possesses three distinct phases through the competition between $|c_2|$ and $q$~\cite{ZZhang2013,XLuo2017}.
For $q \gg 2|c_2|$, the ground state is polar state with all atoms in $\ket{1,0}$.
For $q \ll -2|c_2|$, the ground state becomes TF state with atoms equally populated in $\ket{1,-1}$ and $\ket{1,1}$.
When $-2|c_2| < q < 2|c_2|$, the ground state corresponds to a superposition of all three components.
The two QPT points locate at $q_{c}=\pm|c_2|$.
In this step, we ramp $q(t)$ from $q \gg 2|c_2|$ towards $q \ll -2|c_2|$ with $q(t) = q_0 - \beta t$ to generate the state $\ket{\Psi}_1=e^{-i \int_{0}^{\tau_1} \hat{H}_{\textrm{QPT}}(t) dt}\ket{\Psi}_{\text{in}}$.
Here, $\beta$ denotes the sweeping rate, the TF state can be adiabatically prepared when the sweeping rate is very slow.
Then, a $\pi/2$ pulse with frequency $\omega_L$ is applied on the state $\ket{\Psi}_1$ to generate the input state $\ket{\Psi}_2=\hat{\textrm{R}}_{\pi/2}\ket{\Psi}_1$, as shown in Fig.~\ref{Fig1}(a).
The frequency $\omega_L$ is slightly detuned from the atomic transition frequency $\omega$.
For simplicity, we assume the $\frac{\pi}{2}$ pulse is perfect.
In the interrogation step, the system goes through a phase accumulation process and the output state is $\ket{\Psi({\delta})}_3=\hat{U}({\delta})\ket{\Psi}_2$, as shown in Fig.~\ref{Fig1}(b).
In the recombination step, a second $\frac{\pi}{2}$ pulse(using the same laser) is applied to generate state $\ket{\Psi({\delta})}_4=\hat{R}_{\frac{\pi}{2}}^{\dag}\ket{\Psi({\delta})}_3$, and then ramping $q(t)$ from $q \ll -2|c_2|$ towards $q \gg 2|c_2|$ with $q(t) = q_f + \beta t$, as shown in Fig.~\ref{Fig1}(c).
Thus, the final state after the total sequence can be written as
\begin{equation}\label{Eq:state_f}
\ket{\Psi({\delta})}_{\text{f}}\!=\!e^{-i\! \int_{\tau_1+T}^{\tau_2}\!\! \hat{H}_{\textrm{QPT}}(t) dt} \hat{R}_{\frac{\pi}{2}}^{\dag} \hat{U}({\delta})\hat{R}_{\frac{\pi}{2}} e^{-i\! \int_{0}^{\tau_1}\!\! \hat{H}_{\textrm{QPT}}(t) dt}\ket{\Psi}_{\text{in}}.\\
\end{equation}
Here, $\hat{U}\!(\delta)\!=\!e^{-\frac{i\delta T}{2}(\hat{a}^{\dag}_{1}\hat{a}_{1}\!-\hat{a}_{-1}\hat{a}^{\dag}_{-1})}$ describes the phase accumulation process with $\phi=\delta T$ and $\delta\!=\!\omega\!-\!\omega_{L}$,
$\hat{\textrm{R}}_{\pi/2}\!=\!e^{i\frac{\pi}{4}(\hat{a}^{\dag}_{1}\hat{a}_{-1}+\hat{a}^{\dag}_{-1}\hat{a}_{1})}$ is the $\frac{\pi}{2}$ pulse.
According to Eq.~\eqref{Eq:state_f}, the final state contains the information of the estimated transition frequency $\omega$.
In the measurement step, applying a suitable observable measurement $\hat{O}$ on the final state and the expectation of $\hat{O}$ is
\begin{equation}\label{Eq:Expectation}
\langle\hat{O}({\delta})\rangle=_{\text{f}}\bra{\Psi({\delta})} \hat{O} \ket{\Psi({\delta})}_{\text{f}}.
\end{equation}
In the framework of frequency estimation, if the expectation $\langle\hat{O}({\delta})\rangle$ with respect to detuning $\delta=\omega-\omega_{L}$ is symmetric(or antisymmetric), the expectation $\langle\hat{O}({\delta})\rangle$ can be used to as the frequency lock-in signal and the frequency can be inferred from the pattern of it, as shown in Fig.~\ref{Fig1}(d).
In the next section, we will introduce how to realize the high-precision frequency estimation within this framework.
\section{Frequency measurement\label{Sec3}}
In this section, we illustrate two frequency lock-in signals.
One is the fidelity $F=|_{\text{f}}\langle{\Psi(\delta)}|\Psi\rangle_{\text{in}}|^{2}$ between the final state $\ket{\Psi(\delta)}_{\text{f}}$ and the initial state $\ket{\Psi}_{\text{in}}$.
Another is the expectation of $\hat{N}_0$ on the final state $\ket{\Psi(\delta)}_{\text{f}}$.
Furthermore, we find that the measurement precision of the transition frequency $\omega$ can surpass the SQL and even attain the Heisenberg-limited scaling when the particle number is large enough.
\subsection{Frequency lock-in signals\label{A}}
%
Assuming that the sweeping rate $\beta$ is slow enough and the time-dependent evolutions both are adiabatic in the initialization and recombination steps.
Thus, we have $\ket{\Psi}_1=e^{-i \int_{0}^{\tau_1} \hat{H}_{\textrm{QPT}}(t) dt}\ket{\Psi}_{\text{in}}=\ket{TF}$ and the state after the second $\frac{\pi}{2}$ pulse can be written as
\begin{equation}\label{Eq:state1}
\ket{\Psi(\delta)}_{4}=\hat{\textrm{R}}^{\dagger}_{\pi/2}e^{-\!\frac{i\delta T}{2}(\hat{a}^{\dag}_{1}\hat{a}_{1}-\hat{a}^{\dag}_{-1}\!\hat{a}_{-1})}\hat{\textrm{R}}_{\pi/2}\ket{TF}.
\end{equation}
After some algebra, we can obtain the explicit form of the state $\ket{\Psi(\delta)}_{4}$(see Appendix A for derivation).
For brevity, we denote $n=N/2$ in the follwing.

If $n$ is even, we have $\ket{\Psi(\delta)}_{4}=\ket{\Psi(-\delta)}_{4}$ and it is
\begin{widetext}
\begin{eqnarray}\label{Eq:State-after-BS1-N-even}
\ket{\Psi(\delta)}_{4}
=&&\sum\limits_{m=0}^{n}A({n,m})\ket{2n-2m,0,2m}\nonumber\\
&&+2\sum\limits_{k=0}^{\frac{n}{2}-1}\sum\limits_{m_{1}=0}^{2k}\!\sum\limits_{m_{2}=0}^{2n-2k}B({n,k,m_1,m_2})\cos[(2k-n)\delta T]
\ket{2n-2k+m_{1}-m_{2},0,2k-m_{1}+m_{2}}.
\end{eqnarray}
\end{widetext}
If $n$ is odd, we have $\ket{\Psi(\delta)}_{4}=-\ket{\Psi(-\delta)}_{4}$ and it is
\begin{widetext}
\begin{eqnarray}\label{Eq:State-after-BS1-N-odd}
\ket{\Psi(\delta)}_{4}
=2i\sum\limits_{k=0}^{\frac{n-1}{2}}\sum\limits_{m_1=0}^{2k}\sum\limits_{m_2=0}^{2n-2k}B(n,k,m_1,m_2)\sin[(2k-n)\delta T]\ket{2n-2k+m_1-m_2,0,2k-m_1+m_2}.
\end{eqnarray}
\end{widetext}

Here, the coefficients $A_{n,m}$ and  $B_{n,k,m_1,m_2}$ read as 
\begin{widetext}
\begin{eqnarray}\label{Eq:coeffcient_B}
A_{n,m}=(-1)^{n-m}(\frac{1}{2})^{2n}\frac{C_{n}^{n/2}}{\sqrt{n!n!}}C_{n}^{m}\sqrt{(2m)!(2n-2m)!},\\ \nonumber
\end{eqnarray}
\begin{eqnarray}\label{Eq:coeffcient_A}
B_{n,k,m_1,m_2}=(-1)^{n-k}(\frac{1}{2})^{2n}\sqrt{\frac{C_{2k}^{k}C_{2n-2k}^{n-k}}{2k(2n-2k)!}}
C_{2k}^{m_1}C_{2n-2k}^{m_2}
\sqrt{(2n-2k-m_2+m_1)!(2k-m_1+m_2)!}.
\end{eqnarray}
\end{widetext}
The $C_{i}^{j}$ is the combinatorial number.
Furthermore, the final state is
\begin{equation}\label{Eq:state1}
\ket{\Psi(\delta)}_{\textrm{f}}=e^{-i \int_{\tau_1+T}^{\tau_2} \hat{H}_{\textrm{QPT}}(t) dt} \ket{\Psi(\delta)}_{4}.
\end{equation}
According to Eq.~\eqref{Eq:state1}, we have $\ket{\Psi(\delta)}_{\textrm{f}}\!=\!\ket{\Psi(-\delta)}_{\textrm{f}}$ when $n$ is even, and $\ket{\Psi(\delta)}_{\textrm{f}}\!=-\!\ket{\Psi(-\delta)}_{\textrm{f}}$ when $n$ is odd.
Thus, the fidelity $F(\delta)$ and population measurement $\langle\hat{N}_{0}(\delta)\rangle$ both are symmetric with respect to the lock-in point $\omega=\omega_{L}$, i.e.,  $F(\delta)\!=\!F(-\delta)$ and $\langle\hat{N}_{0}(\delta)\rangle=\langle\hat{N}_{0}(-\delta)\rangle$.
They both can be the frequency lock-in signals to obtain the value of $\omega$.
Especially, when $\delta=0$, the state $\ket{\Psi(\delta=0)}_{4}=\ket{TF}$, and the final state is
\begin{eqnarray}\label{Eq:state}
\ket{\Psi(\delta=0)}_{\textrm{f}}&=&e^{-i \int_{\tau_1+T}^{\tau_2} \hat{H}_{\textrm{QPT}}(t) dt} \ket{\Psi(\delta=0)}_{4}\nonumber \\
&=&e^{-i \int_{\tau_1+T}^{\tau_2} \hat{H}_{\textrm{QPT}}(t) dt}\ket{TF}=\ket{\Psi}_{\text{in}}.
\end{eqnarray}
Thus, we have
\begin{eqnarray}\label{Frequency locking 1}
F(\delta=0)
            =|_{\text{in}}\langle{\Psi}|\Psi\rangle_{\text{in}}|^{2}=1,
\end{eqnarray}
and
\begin{eqnarray}\label{Frequency locking 2}
\langle\hat{N}_0(\delta=0)\rangle
= \bra{0,N,0} {\hat{N}_0}\ket{0,N,0}=N.
\end{eqnarray}
According to Eq.~\eqref{Frequency locking 1} and Eq.~\eqref{Frequency locking 2}, we find that the two frequency lock-in signals approach to their maximum when $\omega-\omega_{L}=0$. 
In Fig.~\ref{Fig2}, the variation of fidelity $F(\delta)$ and population measurement $\langle\hat{N}_{0}(\delta)\rangle$ versus detuning $\delta$ are shown.
The numerical results agree perfectly with our theoretical predictions.
Thus, one can determine the frequency lock-in point $\omega-\omega_{L}=0$ from the pattern of the two frequency lock-in signals.
\subsection{Measurement precision\label{B}}
\begin{figure}[!htp]
 \includegraphics[width=1\columnwidth]{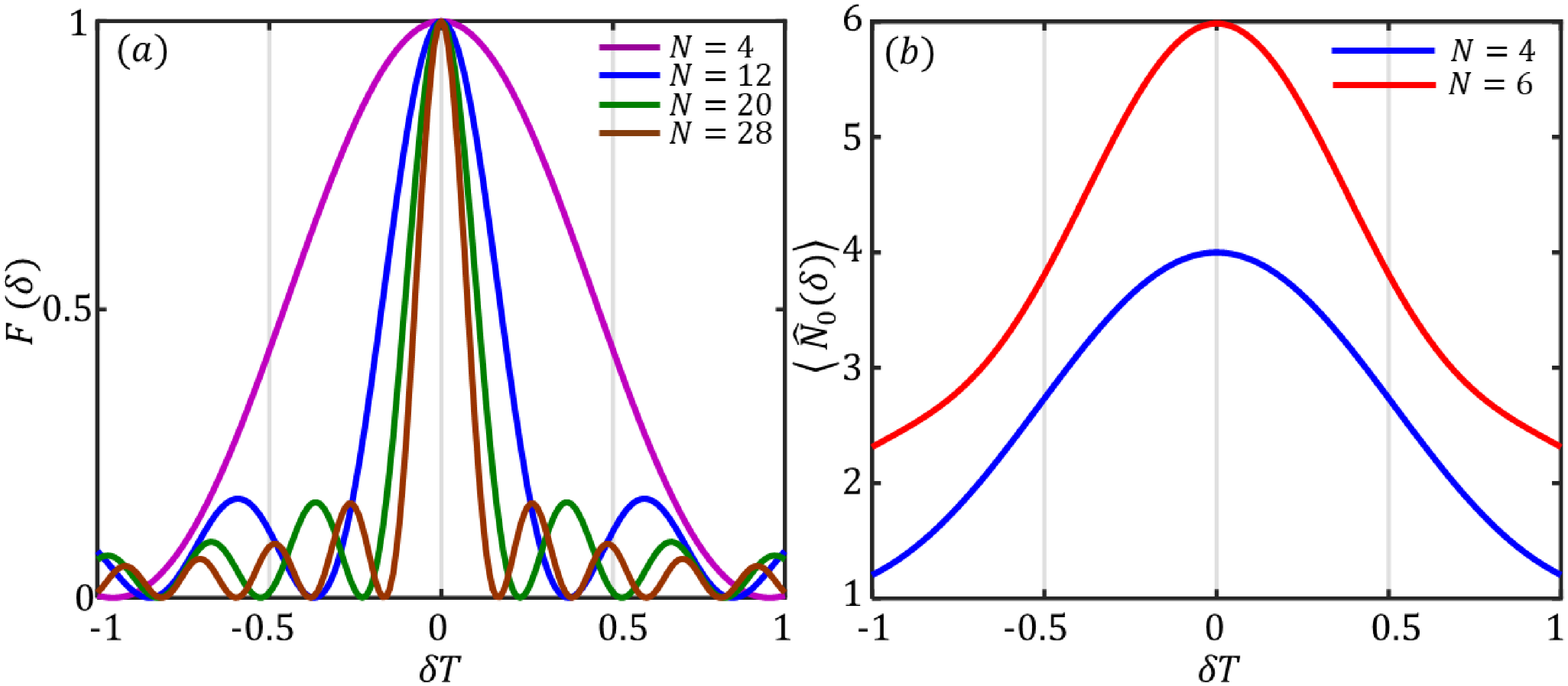}
  \caption{\label{Fig2}(color online).
  The two frequency lock-in signals of our scheme. (a)The variation of frequency lock-in signal $F(\delta)$ versus detuning $\delta$ for different $N$. The expectation of $F(\delta)$ is symmetric with respect
  to the lock-in point $\omega-\omega_{L}=0$.
  (b)The variation of frequency lock-in signal $\langle \hat{N}_0 (\delta) \rangle $ versus detuning $\delta$ for different particle number $N$. The expectation of $\langle \hat{N}_0 (\delta) \rangle $ is symmetric with respect
  to the lock-in point $\omega-\omega_{L}=0$.
  Here, $c_{2}=-1$, $q_0 =-q_f=3$, $T=100$ and $ \beta = 0.01.$}
\end{figure}
In this subsection, we illustrate the measurement precision of frequency within our scheme.
We analyze the measurement precision via two conventional methods, one is according to the linewidth of fidelity $F(\delta)$, another is according to the error propagation formula via population number $\hat{N}_0$ .

\begin{figure}[!htp]
  \includegraphics[width=1\columnwidth]{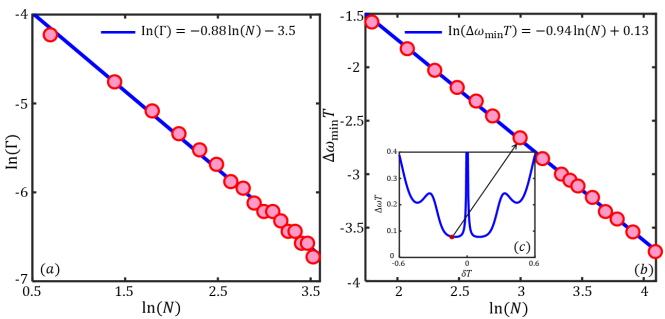}
  \caption{\label{Fig3}(color online).
  (a)Log-log scaling of the linewidth $\Gamma$ versus the total particle number.
  The blue line is the fitting curve.
  (b) Log-log scaling of the optimal measurement precision $\Delta{\omega}_{\textrm{min}}$ versus total particle number.
  The blue line is the fitting curve.
  (c) The variation of measurement precision of frequency $\omega$ versus detuning $\delta$ for $N=20$.
  The black red dot corresponds to the optimal measurement precision $\Delta{\omega}_{\textrm{min}}$, which is near $\delta=0$.
  Here, $c_{2}=-1$, $q_0 =-q_f=3$, $T=100$ and $ \beta = 0.01.$}
\end{figure}

Here, we analyze the measurement precision via the linewidth of fidelity $F(\delta)$.
The value of linewidth is defined as the full width half maximum(FWHM) and we denote it as $\Gamma$.
As shown in Fig.~\ref{Fig2}(a), it is evident that the linewidth $F(\delta)$ is a function of detuning $\delta$ and decreases as $N$ increases.
To confirm the dependence of $\Gamma$ on the total particle number $N$, we numerically calculate the linewidth $\Gamma$ versus $N$, as shown in Fig.~\ref{Fig3}(a).
According to the fitting result, the log-log linewidth is $\textrm{In}({\Gamma})\approx -0.88 \textrm{In}(N)-3.5$.
For $N$ uncorrelated atoms, the log-log linewidth is $\textrm{In}({\Gamma})\approx -0.5 \textrm{In}(N)$, thus our scheme can decrease the linewidth $\Gamma$ effectively.
%
%

%

Further, we consider the measurement precision via population measurement of $\hat{N}_{0}$ on the final state $\ket{\Psi(\delta)}_\textrm{f}$.
According to the error propagation formula, the measurement precision of frequency $\omega$ is
\begin{equation}\label{Eq:Frequency uncertainly}
\Delta \omega=\frac{\Delta{\hat{N}_{0}}}{|\partial{\langle\hat{N}_{0}(\delta)\rangle_f}/ \partial{\omega}|}.
\end{equation}
Here, $\Delta{\hat{N}_{0}}$ is the standard deviation of $\hat{N}_{0}$ and can be written as
\begin{equation}\label{Eq:Deviation}
\Delta{\hat{N}_{0}}=\sqrt{\langle{\hat{N}_{0}^2(\delta)}\rangle-(\langle{\hat{N}_{0}(\delta)}\rangle)^2}.
\end{equation}
In Fig.~\ref{Fig3}(c), how the measurement precision $\omega$ changes with detuning $\delta$ is shown, and we find the optimal measurement precision $\Delta \omega_{\textrm{min}}$ occurs near the frequency lock-in point $\omega-\omega_{L}=0$.
To confirm the dependence of $\Delta \omega$ on the total particle number $N$, we numerically calculate the variation of the measurement precision versus particle number, as shown in Fig.~\ref{Fig3}(b).
According to the fitting result, the optimal log-log measurement precision $\Delta \omega_{\textrm{min}}$ is $\textrm{In}(\Delta \omega_{\textrm{min}})\approx -0.94 \textrm{In}(N)+0.13$.
It indicate that the combination of reversed adiabatic driving and population measurement is an effective way to realize quantum-enhanced frequency estimation with TF state.
%
%
%
%
\section{Robustness against imperfections \label{Sec4}}
%
In this section, we study the robustness of our scheme.
In practical experiments, there are many imperfections that influence the frequency lock-in signal and limit the final measurement precision.
Here, we discuss two imperfections: the detection noise in the measurement step and the non-adiabatic effect in the initialization and recombination steps.
%
\subsection{Robustness against detection noise \label{A}}
\begin{figure}[!htp]
  \includegraphics[width=1\columnwidth]{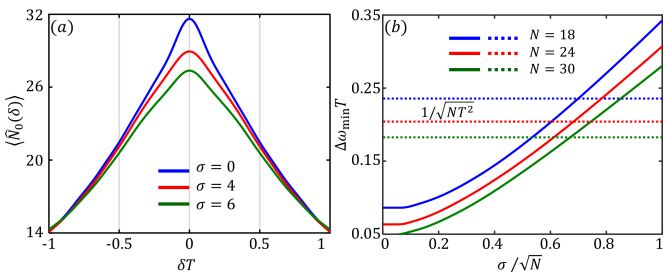}
  \caption{\label{Fig4}(color online).
  The influence of detection noise on frequency estimation.
  (a)The variation of the frequency lock-in signal $\langle {N_{0}(\delta)}\rangle$ versus detuning $\delta$ with different detection noise $\sigma$ under $N=30$.
  (b)The optimal measurement precision $\omega_\textrm{{min}}$  versus detection noise $\sigma$ for $N=18, 24, 30$.
  The dotted line is the SQL, i.e. $1/(\sqrt{N T^{2}})$.
  Here, $c_{2}=-1$, $q_0 =-q_f=3$, $T=100$ and $ \beta = 0.01.$}
\end{figure}
We study the influence of detection noise by considering the additional classical noise in the measurement process.
In an ideal situation, the population  measurement on the final state can be rewritten as $\langle\hat{N}_{0}(\delta)\rangle=\sum_{N_0}{P(N_0|\delta)}N_0$, where ${P(N_0|\delta)}$ is the ideal conditional probability which obtain measurement result $N_0$ with a given $\delta$.
However, in realistic experiment, the detection noise can limit the measurement precision of frequency.
For an imperfect detector with Gaussian detection noise~\cite{LucaPezze2013,DMStamperKurn2013,Gabbrielli163002,SPNolan2017}, the population measurement becomes
\begin{equation}\label{Eq:detection_noise}
\langle\hat{N}_{0}(\delta)\rangle=\sum_{N_0}\tilde{P}(N_0|{\sigma})N_0,
\end{equation}
with
\begin{equation}\label{Eq:detection_noise_p}
{\tilde{P}(N_0|{\sigma})}=\sum_{\tilde{N}_0} A_{\tilde{N}_0}{e^{-(N_0-\tilde{N}_0)^2/2{\sigma}^2}}{P({N}_0|{\delta})}.
\end{equation}
the conditional probability depends on the detection noise $\sigma$.
Here, $A_{\tilde{N}_0}$ is a normalization factor.
%
%

According to our numerical calculation, we find that the Gaussian detection noise does not induce any frequency shift on the frequency lock-in signal $\langle\hat{N}_0(\delta)\rangle$.
The frequency lock-in signal $\langle\hat{N}_0(\delta)\rangle$ is still symmetric with respect to the lock-in point $\omega=\omega_{L}$ and attains its maximum at the lock-in point.
However, the height and the sharpness of the peak both decrease as $\sigma$ increases, as shown in Fig.~\ref{Fig4}(a).
To further confirm the influence of the detection noise, the optimal measurement precision $\Delta{\omega}_{\textrm{min}}$ versus the detection noise $\sigma$ for different particle number $N$ is shown in Fig.~\ref{Fig4}(b).
The measurement precision can still beat the SQL when $\sigma\leq 0.7 \sqrt{N}$ with $N\in[2,32]$.
Thus our proposal is robust against to detection noise.
\subsection{Influences of non-adiabatic effect\label{B}}
To realize perfect initialization and recombination, the sweeping process should be adiabatic.
However, non-adiabatic effect always exists in practical experiments.
In our scheme, non-adiabatic effect can influence the initialization and recombination steps.
In general, the adiabaticity of the driving process can be characterized by the sweeping rate $\beta$.
If the sweeping rate $\beta$ is sufficiently small, the adiabatic evolution can be achieved.

\begin{figure}[!htp]
  \includegraphics[width=1\columnwidth]{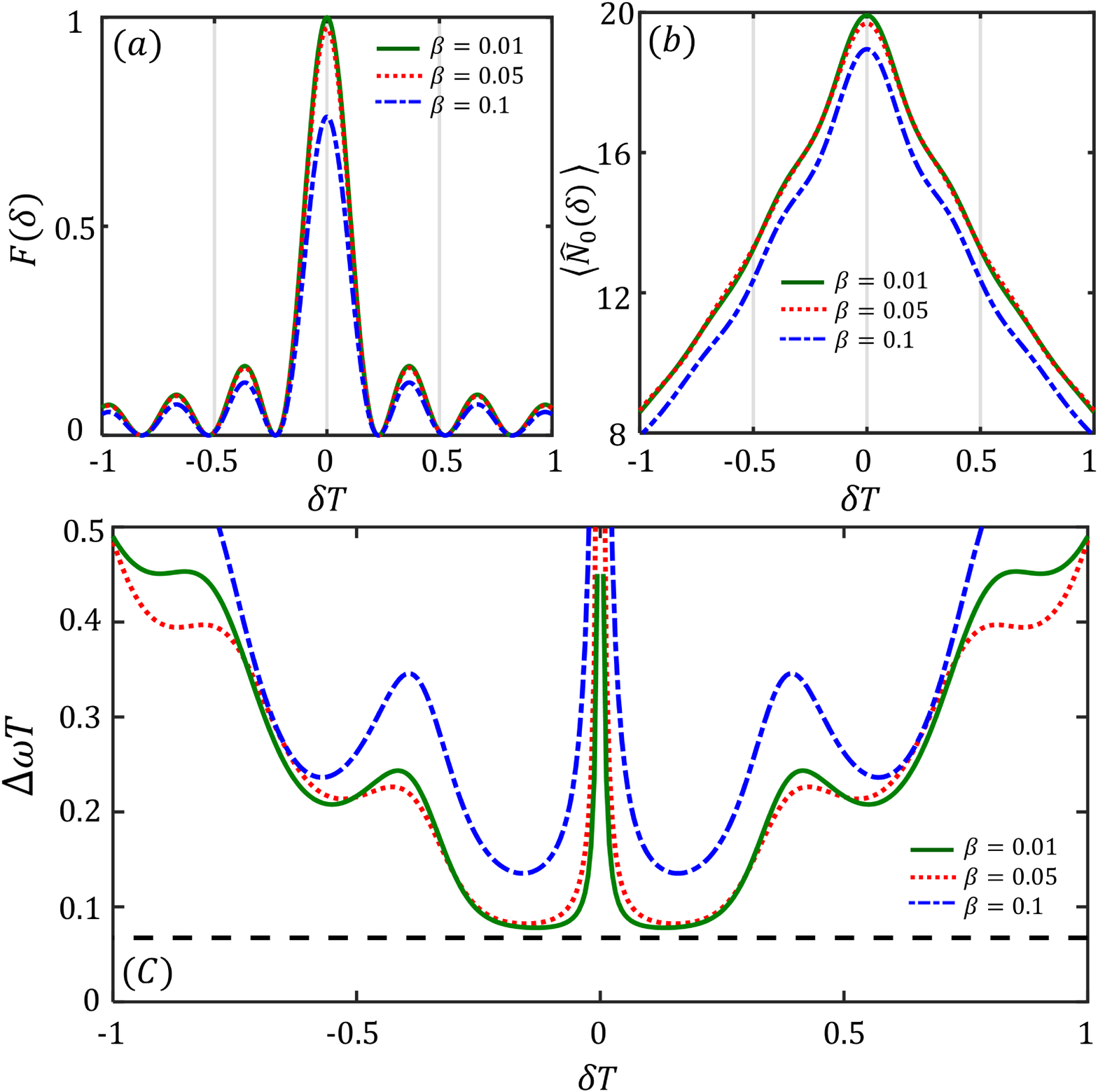}
  \caption{\label{Fig5}(color online).
  The influences of non-adiabatic effect on frequency estimation.
   (a)The variation of frequency lock-in signal $F(\delta)$ with detuning $\delta$ for different sweeping rate $\beta$ under $N=30$;
   (b)The variation of frequency lock-in signal $\langle {\hat{N}_{0}(\delta)}\rangle$ with detuning $\delta$ for different sweeping rate $\beta$;
   (c)The measurement precision $\Delta \omega$ versus detuning $\delta$ for different sweeping rate $\beta$ under $N=20$.
   Here, $c_{2}=-1$, $q_0 =-q_f=3$, $T=100$ and $\beta = 0.01.$}
\end{figure}

To confirm the influences of non-adiabatic effect on frequency estimation, the variations of the two frequency lock-in signals with different sweeping rate $\beta$ are shown in Figs.~\ref{Fig5}(a) and (b).
Our results indicate that the non-adiabatic effect also does not induce any frequency shift on the two frequency lock-in signals.
The height and the sharpness of the peak also decrease as $\beta$ increases.
Further, we study the influence of non-adiabatic effect on the measurement precision, the variation of $\Delta\omega$ with detuning $\delta$ for different sweeping rate $\beta$ is shown in Fig.~\ref{Fig5}(c).
It can be observed that the measurement precision of frequency $\Delta\omega$ becomes worse as $\beta$ increases, but the measurement precision of frequency $\Delta\omega$ can still beat the SQL when the sweeping rate $\beta$ is moderate.
%
\section{summary \label{Sec5}}
In summary, we have presented a realizable scheme for performing Heisenberg-limited frequency estimation with spin-$1$ Bose-Einstein condensate by driving through QPTs.
In our scheme, by adjusting the laser frequency $\omega_{L}$, we can extract the transition frequency $\omega$ according to the frequency lock-in signals.
For every fixed laser frequency $\omega_{L}$, one can implement quantum interferometry for frequency measurement.
The interferometry consists of four steps: initialization, interrogation, recombination and measurement.
In our scheme, we combine adiabatically driving through QPTs with $\frac{\pi}{2}$ pulse to realize the initialization and recombination.
Based upon the proposed scheme, we find two frequency lock-in signals: fidelity and population measurement.
They are both symmetric with respect to lock-in point $\omega=\omega_{L}$ and achieve their maximum at the lock-in point.
Thus, one can obtain the value of $\omega$ from the two frequency lock-in signals.
Further, we study the measurement precision of frequency via two different methods.
We find the measurement precision $\Delta\omega$ can exhibit Heisenberg-limited scaling via population measurement.

At last, we illustrate the robustness of our scheme against detection noise and non-adiabatic effect.
These imperfections do not induce any frequency shift on the frequency lock-in signals and just degrade the measurement precision.
For detection noise, the measurement precision $\Delta{\omega}$ can still beat the SQL when $\sigma \leq 0.7 \sqrt{N}$ in our calculation.
For non-adiabatic effect, the measurement precision $\Delta{\omega}$ can still beat the SQL when the sweeping rate $\beta$ is moderate.

In experiment, the TF state has been generated in spin-$1$ atomic Bose-Einstein condensate via driving the system through QPTs ~\cite{ZZhang2013,XLuo2017}.
Meanwhile, the precise implementation of $\pi/2$ pulses also is a mature technology in quantum control.
Thus, our scheme adds no additional complexity for the apparatus design.
Compared with the conventional quantum-enhanced frequency estimation with entanglement, our proposal does not require single-particle resolution detectors and is robust against detection noise~\cite{PRA023810,JJBollinger1996}.
%
%
Our study paves a new way to realize Heisenberg-limited frequency measurement.

\acknowledgements{This work is supported by the National Natural Science Foundation of China (12025509, 11874434), the Key-Area Research and Development Program of GuangDong Province (2019B030330001), and the Science and Technology Program of Guangzhou (201904020024). M. Z. is partially supported by the National Natural Science Foundation of China (12047563). J. H. is partially supported by the Guangzhou Science and Technology Projects (202002030459). }

\setcounter{equation}{0}
\renewcommand{\theequation}{A\arabic{equation}}
%
%
\section*{APPENDIX A: the effect of $\frac{\pi}{2}$ pulse on the general Fock state}
Here, we give the derivation of  Eq.\eqref{Eq:State-after-BS1-N-odd} and Eq.\eqref{Eq:State-after-BS1-N-even} in the main text.
The $\frac{\pi}{2}$ pulse operation $\hat{\textrm{R}}_{\pi/2}=e^{-i\frac{\pi}{4}(\hat{a}^{\dag}_{1}\hat{a}_{-1}-\hat{a}^{\dag}_{-1}\hat{a}_{1})}$ just acts on $m=\!\pm1$ modes.
For convenience, we rewrite the $\hat{\textrm{R}}_{\pi/2}^{\dagger}$ to $\hat{\textrm{U}}_{\pi/2}=e^{-i\frac{\pi}{4}(\hat{a}^{\dag}\hat{b}-\hat{b}^{\dag}\hat{a})}$.
Here, $\hat{a}$ and $\hat{b}$ are the annihilation operators for particles in mode $\ket{a}$ and mode $\ket{b}$, respectively.
For a two mode Fock state $\ket{n_{a},n_{b}}$, we give the general derivation of $\hat{U}_{\theta}\ket{n_{a},n_{b}}$ with $\hat{U}_{\theta}=e^{-i\theta(\hat{a}^{\dag}\hat{b}-\hat{b}^{\dag}\hat{a})}$,
\begin{eqnarray}\label{Eq:pulse1}
\hat{U}_{\theta}\ket{n_{a},n_{b}}&=&\hat{U}_{\theta}\frac{(\hat{a}^{\dagger})^{n_{a}}}{\sqrt{n_a!}}\frac{(\hat{b}^{\dagger})^{n_{b}}}{\sqrt{n_b!}}
\ket{0}_a\ket{0}_b\nonumber \\
&=&\hat{U}_{\theta}\frac{(\hat{a}^{\dagger})^{n_{a}}}{\sqrt{n_a!}}\frac{(\hat{b}^{\dagger})^{n_{b}}}{\sqrt{n_b!}}\hat{U}_{\theta}^{\dagger}\ket{0}_a\ket{0}_b.
\end{eqnarray}
Due to $\hat{U}_{\theta}\hat{U}_{\theta}^{\dagger}=1$, we have
\begin{eqnarray}\label{Eq:pulse2}
\hat{U}_{\theta}\ket{n_{a},n_{b}}=\frac{(\hat{U}_{\theta}\hat{a}^{\dagger}\hat{U}_{\theta}^{\dagger})^{n_{a}}}{\sqrt{n_a!}}\frac{(\hat{U}_{\theta}\hat{b}^{\dagger}\hat{U}_{\theta}^{\dagger})^{n_{b}}}{\sqrt{n_b!}}\ket{0}_a\ket{0}_b \nonumber. \\
\end{eqnarray}
Then, using the basic formula: $e^{\hat{A}}\!\hat{B}e^{-\!\hat{A}}=\sum_{m}^{\infty}\frac{1}{m!}[\hat{A}^{\!(i)},\!\hat{B}]$ and the Taylor expansion, we have

\begin{eqnarray}\label{Eq:pulse3}
\hat{U}_{\theta}\hat{b}^{\dagger}\hat{U}_{\theta}^{\dagger}=\textrm{cos}{\theta}\hat{b}^{\dagger}+ \textrm{sin}{\theta}\hat{a}^{\dagger},\\
\hat{U}_{\theta}\hat{a}^{\dagger}\hat{U}_{\theta}^{\dagger}=\textrm{cos}{\theta}\hat{a}^{\dagger}- \textrm{sin}{\theta}\hat{b}^{\dagger}.
\end{eqnarray}
Similarly, we can obtain
\begin{eqnarray}\label{Eq:pulse4}
\hat{U}_{\theta}\hat{b}\hat{U}_{\theta}^{\dagger}= \textrm{cos}{\theta}\hat{b}+ \textrm{sin}{\theta}\hat{a},  \\
\hat{U}_{\theta}\hat{a}\hat{U}_{\theta}^{\dagger}= \textrm{cos}{\theta}\hat{a}- \textrm{sin}{\theta}\hat{b}.
\end{eqnarray}
When $\theta=\pi/4$, we have
\begin{eqnarray}\label{Eq:pi/2 pulse}
\frac{(\hat{U}_{\pi/2}\hat{a}^{\dagger}\hat{U}_{\pi/2}^{\dagger})^{n_{a}}}{\sqrt{n_a!}}\frac{(\hat{U}_{\pi/2}\hat{b}^{\dagger}\hat{U}_{\pi/2}^{\dagger})^{n_{b}}}{\sqrt{n_b!}}\ket{0}_a\ket{0}_b \nonumber \\
=\frac{\sqrt{2}(\hat{a}^{\dagger}-\hat{b}^{\dagger})^{n_a}}{{2}\sqrt{n_a!}}
\frac{\sqrt{2}(\hat{a}^{\dagger}+\hat{b}^{\dagger})^{n_{b}}}{{2}\sqrt{n_b!}}\ket{0}_a\ket{0}_b.
\end{eqnarray}
Especially, when $n_a=n_b=n=N/2$, we have
\begin{eqnarray}\label{Eq:pi/2 pulse6}
&&\frac{(\hat{U}_{\pi/2}\hat{a}^{\dagger}\hat{U}_{\pi/2}^{\dagger})^{n}}{\sqrt{n!}}\frac{(\hat{U}_{\pi/2}\hat{b}^{\dagger}\hat{U}_{\pi/2}^{\dagger})^{n}}{\sqrt{n!}}\ket{0}_a\ket{0}_b \\ \nonumber
&=&\frac{1}{n!2^{n}}({\hat{a}^{\dagger}\hat{a}^{\dagger}-\hat{b}^{\dagger}\hat{b}^{\dagger}})^{n}\ket{0}_a\ket{0}_b \\ \nonumber
&=&\sum_{k=0}^{n}D_n^{k}\ket{2k}_a\ket{2n-2k}_b.
\end{eqnarray}
Here, $D_n^{k}=(-1)^{(n-k)}\frac{\sqrt{2k!(2n-2k)!}{n!}}{k!(n-k)!}$.
Thus, the Eq.\eqref{Eq:state1} in the main text can be written as
\begin{eqnarray}\label{Eq:pi/2 pulse7}
&&\ket{\Psi(\delta)}=\hat{\textrm{R}}^{\dagger}_{\pi/2}e^{-\frac{i\delta T}{2}(\hat{a}^{\dag}_{1}\hat{a}_{1}-\hat{a}^{\dag}_{-1}\hat{a}_{-1})}\textrm{R}_{\pi/2}\ket{\Psi}_{TF}\nonumber \\
&=&e^{-{i\delta(2k-n)T}}\sum_{k=0}^{n}C_n^{k}\hat{\textrm{R}}^{\dagger}_{\pi/2}\ket{2k,0,2n-2k}\nonumber \\
&=& \!e^{-{i\delta(2k-n)T}} \sum_{k=0}^{n}C_n^{k}\frac{[\hat{\textrm{R}}^{\dagger}_{\pi/2}\hat{a}^{\dagger}_1\hat{\textrm{R}}_{\pi/2}]^{2k}[\hat{\textrm{R}}^{\dagger}_{\pi/2}\hat{a}^{\dagger}_{-1}\hat{\textrm{R}}_{\pi/2}]^{2n\!-\!2k}}{\sqrt{2k!(2n-2k)!}}\ket{0,0,0}.\nonumber\\
\end{eqnarray}
Then, submitting the Eq.\eqref{Eq:pulse3} and Eq.\eqref{Eq:pulse4} into Eq.\eqref{Eq:pi/2 pulse7}, we can obtain Eq.\eqref{Eq:State-after-BS1-N-odd} and Eq.\eqref{Eq:State-after-BS1-N-even} straightly in the main text.
Such as, for $n=2$, the state $\ket{\psi(\delta)}$ can be written as
\begin{eqnarray}\label{Eq:State-after-BS1-N-4}
\ket{\psi(\delta)}&=&(\frac{3}{4}\textrm{cos}(4\delta)+\frac{1}{4})\ket{2,0,2}\nonumber \\
&+&i{\frac{\sqrt {3}}{2\sqrt {2}}}\textrm{sin}(4\delta)[\ket{1,0,3}+\ket{3,0,1}]\nonumber \\
&+&{\frac{\sqrt {3}}{4\sqrt {2}}}(\textrm{cos}(4\delta)-1)[\ket{4,0,0}+\ket{0,0,4}].
\end{eqnarray}
When $\omega=\omega_{L}$, we have $\ket{\psi(\delta=0)}=\ket{2,0,2}$, and $F(\delta=0)=1$.

\end{document}